\documentstyle[art12]{article}
\def \sect #1 {\setcounter{equation} 0\section{#1}}
\def \be  {\begin{equation}}
\def \ee  {\end{equation}}
\def \ba  {\begin{eqnarray}}
\def \ea  {\end{eqnarray}}
\def \baa {\begin{eqnarray*}}
\def \eaa {\end{eqnarray*}}
\def \bb  {\begin{thebibliography}}
\def \eb  {\end{thebibliography}}
\newcommand \ci [1] {\cite{#1}}
\newcommand \bi [1] {\bibitem{#1}}
\def \lab #1 {\label{#1}}
\newcommand\re[1]{(\ref{#1})}
\def \qqquad {\qquad\quad}
\def \qqqquad {\qquad\qquad}
\newcommand\lr[1]{{\left({#1}\right)}}

\def \tr {\mbox{tr}}

\newcommand \ket [1] {|{#1}\rangle}

\def \e {\mbox{e}}

\def \Ch {{\tt h}}
\def \CH {{\cal H}}

\font\cmss=cmss10 \font\cmsss=cmss10 at 7pt
\def\inbar{\,\vrule height1.5ex width.4pt depth0pt}
\def\IC{\relax\hbox{$\inbar\kern-.3em{\rm C}$}}
\def\IZ{\relax\ifmmode\mathchoice
{\hbox{\cmss Z\kern-.4em Z}}{\hbox{\cmss Z\kern-.4em Z}}
{\lower.9pt\hbox{\cmsss Z\kern-.4em Z}}
{\lower1.2pt\hbox{\cmsss Z\kern-.4em Z}}\else{\cmss Z\kern-.4em Z}\fi}
\def\IR{{\hbox{{\rm I}\kern-.2em\hbox{\rm R}}}}

\newcommand \partder [1] {{\partial \over\partial #1}}

\newcommand \bin [2] {{\left({#1}\atop{#2}\right)}}
\def \as {\relax\ifmmode\alpha_s\else{$\alpha_s${ }}\fi}

\def \al #1 {\frac {\as({#1})}{\pi} }
\def \ds #1 {\ooalign{$\hfil/\hfil$\crcr$#1$}}

\begin{document}

\def\thefootnote{\fnsymbol{footnote}}
\thispagestyle{empty}
\hfill\parbox{35mm}{{\sc ITP--SB--94--14}\par
                         hep-th/9404173  \par
                         April, 1994}
\vspace*{45mm}
\begin{center}
{\LARGE High energy QCD as a completely integrable model}
\par\vspace*{22mm}\par
{\large L.~D.~Faddeev}$\ {}^{1,2}$ and  {\large G.~P.~Korchemsky}$\ {}^{3,}$%
\footnote{On leave from the Laboratory of Theoretical Physics,
          JINR, Dubna, Russia}
\par\bigskip\par\medskip
${}^1${\em St.Petersburg Branch of Steklov Mathematical Institute,\par
Fontanka 27, St. Petersburg 191011,
Russia} \par\medskip
${}^2${\em Research Institute for Theoretical Physics,
University of Helsinki,\par
Siltavuorenpenger 20C,
SF 00014 Helsinki,
Finland}
\par\medskip
${}^3${\em Institute for Theoretical Physics, \par
State University of New York at Stony Brook, \par
Stony Brook, New York 11794 -- 3840, U.S.A.}
\end{center}
\vspace*{20mm}

\begin{abstract}
We show that the one-dimensional lattice model proposed by Lipatov
to describe the high energy scattering of hadrons in multicolor QCD
is completely integrable. We identify this model as the XXX Heisenberg
chain of noncompact spin $s=0$ and find the conservation laws of the
model. A generalized Bethe ansatz is developed for the diagonalization
of the hamiltonian and for the calculation of hadron-hadron scattering
amplitude.
\end{abstract}
\newpage
\def\thefootnote{\arabic{footnote}}
\setcounter{footnote} 0


\section{Introduction}

Recently it was suggested by Lipatov \ci{L94} that the asymptotic behavior
of the hadron-hadron scattering amplitudes in QCD in the limit of
large invariant energy $s$ and fixed transferred momentum $t$ can
be described by means of the quantum inverse scattering method \ci{ABA,KIB}.
In the generalized leading logarithmic approximation \ci{L89,Cheng,book2}
the scattering
amplitude is given by the contributions of the Feynman diagrams having
a conserved number $n$ of reggeized gluons in the $t-$channel \ci{Bar,KP}.
For fixed $n$ the scattering amplitude can be
decomposed into $t-$channel partial waves $f_\omega$, describing
$n$ gluon to $n$ gluon scattering with total angular momentum $j=1+\omega$.
The partial wave satisfies the Bethe-Salpeter equation
whose solutions
can be expressed in terms of the wave functions $\chi$ of the compound
states of $n$ gluons as follows \ci{L89,L90}:
$$
f_\omega(\{\vec b_i\},\{\vec b_i^\prime\})=\sum_\alpha \int d^2 b_0
\ \chi_\alpha(\{\vec b_i\};\vec b_0)
\chi^*_\alpha (\{\vec b_i'\};\vec b_0)
$$
Here, $\chi_\alpha(\{\vec b_i\};\vec b_0)$ is the wave function of the
compound state of $n$ reggeized gluons with transverse coordinates
$\{\vec b_i\}=\vec b_1$, $\ldots$, $\vec b_n$ and it is parameterized by
quantum
numbers $\alpha$ and by two-dimensional vector $\vec b_0$ which
represents the center of mass coordinate of the states.
In the special case $n=2$, the function
$\chi(\vec b_1,\vec b_2;\vec b_0)$ is the wave function of the BFKL
pomeron \ci{BFKL,L86}.
Let us define for all impact vectors $\vec b_i=(x_i,y_i)$,
$i=1,\ldots,n$, the following complex
coordinates:
$z_i  = x_i + i y_i$ and
$\bar z_i= x_i - i y_i$
and similarly for the vectors $\vec b_i'$.

Then, in the large $N$ limit, for an arbitrary number $n$ of gluons,
the wave functions $\varphi$ are the eigenstates of the holomorphic
and antiholomorphic hamiltonians \ci{L94,L93,L90}
\be
H_n \chi(\{z_i,\bar z_i\};z_0,\bar z_0)
=\varepsilon_n\chi(\{z_i,\bar z_i\};z_0,\bar z_0),
\qquad
\bar H_n \chi(\{z_i,\bar z_i\};z_0,\bar z_0)
=\bar\varepsilon_n\chi_\omega(\{z_i,\bar z_i\};z_0,\bar z_0),
\lab{H-N}
\ee
with eigenvalues related to complex angular momentum via
$\omega=-\frac{\as}{4\pi} N (\varepsilon_n + \bar\varepsilon_n)$.
The hamiltonians $H_n$ and $\bar H_n$ describe nearest neighbors interaction
of $n$ particles
\be
H_n = \sum_{k=1}^n H_{k,k+1},\qqqquad
\bar H_n = \sum_{k=1}^n \bar H_{k,k+1}
\lab{H-gen}
\ee
with periodic boundary conditions $H_{n,n+1}=H_{n,1}$, $\bar H_{n,n+1}=
\bar H_{n,1}$ and the two-particle hamiltonians are giving by the
equivalent representations
\ba
H_{j,k}&=&P_j^{-1}\log(z_j-z_k)P_j
           +P_k^{-1}\log(z_j-z_k)P_k
           +\log(P_jP_k)+2\gamma_E
\nonumber
\\[3mm]
          &=&2\log(z_j-z_k)+(z_j-z_k)\log(P_jP_k)(z_j-z_k)^{-1}+2\gamma_E
\lab{H1}
\ea
where $P_j=i\partder{z_j}$ and $\gamma_E$ is the Euler constant. The same
operator can also be represented as \ci{L93}
\be
H_{i,k}=\sum_{l=0}^\infty \frac{2l+1}{l(l+1)-\vec L_{ik}^2}-\frac2{l+1}
\,,\qquad \vec L_{ik}^2=-(z_i-z_k)^2\partial_i\partial_k
\lab{H2}
\ee
Thus, in the large $N$ limit the original problem of the calculation of
partial waves is reduced to the diagonalization of two hamiltonians
corresponding to spin chain models defined on a one-dimensional lattice
with the number of sites equal to the number of reggeized gluons in the
$t-$channel.
However, there are additional constraints on the eigenstates of the
QCD hamiltonians imposed by the condition that the partial wave $f_\omega$
is invariant under holomorphic and antiholomorphic
conformal transformations of the coordinates $z$ and $\bar z$ \ci{L86}.
In terms of the wave functions $\chi$
which depend only on parts of the coordinates, this condition leads to
the following transformation properties
\be
\chi(\{z_i,\bar z_i\};z_0,\bar z_0)
\to
\chi(\{z_i',\bar z_i'\};z_0',\bar z_0')
=(c z_0 + d)^{2h} (\bar c \bar z_0 + \bar d)^{2\bar h}
\chi(\{z_i,\bar z_i\};z_0,\bar z_0)
\lab{con1}
\ee
under conformal transformations $z\to z'=\frac{az+b}{cz+d}$ and
$\bar z\to\bar z'=\frac{\bar a \bar z_i+\bar b}{\bar c \bar z_i+\bar d}$
with $ad-bc=\bar a\bar d-\bar b\bar c=1$.
The possible values of the conformal
weights $h$ and $\bar h$ can be parameterized as follows \ci{L86,L89,L90}
\be
     h=\frac{1+m}{2} - i\nu \,, \qqqquad
\bar h=\frac{1-m}{2} - i\nu
\lab{h}
\ee
for arbitrary real $\nu$ and integer $m$.

It was suggested
\ci{L94} that these models can be exactly solved by means of the Bethe ansatz.
Among other things, this would imply that the model contains a
family of ``hidden'' integrals of motion. Indeed, at least one
nontrivial integral of motion $A$ was found \ci{L94}. It commutes with the
hamiltonian and is given by
\be
A=z_{12}z_{23}\cdots z_{n1} P_1 P_2 \cdots P_n \,,\qqqquad
[H_n,A]=0
\lab{A}
\ee
together with the
analogous operator for the antiholomorphic system. It was noticed
\ci{L94}, that
this operator belongs to the family of $n$ mutually commuting operators
generated by a monodromy matrix which satisfies the Yang-Baxter equation.
However it was not clear that all these operators
commute with the hamiltonian and that they are
the integrals of motion of the model.
In the present paper we show that this is indeed the case. We
prove that the model \re{H-N} is completely integrable and
discuss the possibility of solving it by means of the Bethe ansatz.

\section{High energy multicolor QCD as XXX Heisenberg magnet}

Let us show that the hamiltonians \re{H1} and \re{H2} coincide with
that of the XXX Heisenberg magnet for spin $s=0$.
To give meaning to the limit $s=0$ we recall that the quantum
inverse scattering method allows one \ci{KI} to construct a family of
exactly solvable XXX models for arbitrary complex values
of the spin $s$. For arbitrary complex $s$ the spin operators
$S_k^\alpha$ $(\alpha=1,2,3)$ in all sites $k=1,\ldots,n$ of the
lattice can be realized as follows
\be
S_k^+ = z_k^2\partial_k-2s z_k ,
\qquad
S_k^- = -\partial_k ,
\qquad
S_k^3 = z_k \partial_k - s
\lab{S}
\ee
The definition of the model
is based on the existence of a fundamental matrix $R_{f_1f_2}(\lambda)$
which obeys the Yang-Baxter equation and acts in the auxiliary space
$V_{f_1}\otimes V_{f_2}=\Ch\otimes\Ch$ with the space $\Ch$ having
the dimension of the local quantum space in each site.
For arbitrary complex spin $s$ it is given by
the following expression \ci{TTF}
\be
R_{f_1f_2}(\lambda)=f(s,\lambda)
\frac{\Gamma(i\lambda-2s)\Gamma(i\lambda+2s+1)}
                         {\Gamma(i\lambda-J)\Gamma(i\lambda+J+1)}
\lab{Rf}
\ee
where the operator $J$ is defined in the space $V_{f_1}\otimes V_{f_2}$
as a solution of the operator equation
$$
J(J+1) = 2\vec S\otimes\vec S +2s(s+1)
$$
Notice that the definition of the $R-$matrix contains ambiguity
in multiplication by an arbitrary c-number valued function $f(s,\lambda)$
which is fixed by imposing additional normalization conditions on
$R_{f_1f_2}(\lambda)$. For the special value $\lambda=0$ of the
spectral parameter the operator $R_{f_1f_2}(0)$ is proportional to the
permutation operator $P_{f_1f_2}$ on $V_{f_1}\otimes V_{f_2}$ and
both operators are equal to each other for finite dimensional representations
of the spin operators \ci{TTF} provided that $f(s,0)=1$.
The hamiltonian of
the XXX model of spin $s$ is given
by the general expression \re{H-gen} with the two-particle hamiltonian defined
as
\be
H_{12}=\left.\frac1{i}\frac{d}{d\lambda}
    \log R_{f_1f_2}(\lambda)\right\vert_{{}_{\lambda=0}}
\lab{H12}
\ee
Let us consider the expression for the hamiltonian $H$ in the limit
$s=0$. Using the expressions \re{Rf} and \re{H12} we find that
one of the $\Gamma-$functions in the numerator of \re{Rf} diverges
as $\lambda,s\to 0$. However we use the possibility to choose
$f(s,\lambda)=\Gamma(i\lambda+2s+1)/\Gamma(i\lambda-s)$
to avoid this problem and obtain the two-particle hamiltonian as
\be
H_{ik}=-\psi(-J_{ik})-\psi(J_{ik}+1)+2\psi(1)
\lab{psi}
\ee
Here $\psi(x)=d \log\Gamma(x)/dx$ and the operator $J_{ik}$ is one of the
solutions of the equation
$$
J_{ik}(J_{ik}+1)=2\vec S_i\otimes\vec S_k=-(z_i-z_k)^2\partial_i\partial_k
$$
where we substituted the explicit form \re{S} of the spin operators for $s=0$.
Comparing the last relation with \re{H2} we notice that the operator
$J_{ik}(J_{ik}+1)$
coincides with the definition of the operator $L_{jk}^2$.
Substituting  $L_{jk}^2=J_{jk}(J_{jk}+1)$ into \re{H2} one can perform the
summation over $l$ to get
an expression for the hamiltonian which is identical to that in \re{psi}!
Thus, we identify the holomorphic and antiholomorphic QCD hamiltonians
\re{H-N} as hamiltonians of a one-dimensional XXX Heisenberg model with spin
$s=0$. In what follows, we study only the holomorphic sector of the model.
The generalization to the antiholomorphic sector is straightforward.

The identification of the model as XXX magnet means that the system
\re{H-N}, which describes high-energy asymptotics in multi-color QCD,
is exactly
solvable. To find the family of local integrals of motion of the model we
follow the standard procedure. To any site $k$ we assign auxiliary and
fundamental Lax operators \ci{TTF}
\be
L_{k,a}(\lambda)=\lambda I_n\otimes I_a + i \vec S_k \otimes \vec\sigma_a
\,,\qquad
L_{k,f}(\lambda)=R_{k,f}(\lambda)
\lab{Lax}
\ee
where $\vec \sigma$ are the Pauli matrices. The operators $L_{k,a}$
and $L_{k,f}$ act locally in the space $\Ch_k\otimes \IC^2$ and
$\Ch_k\otimes\Ch$, respectively, where $\Ch_k$ is the quantum space in the
$k-$th site and the dimensions of $\Ch$ and $\Ch_k$ coincide.
Taking the ordered product of the Lax operators along the lattice we define
the auxiliary monodromy matrix as a matrix in $\IC^2$
\be
T_a(\lambda) = L_{n,a}(\lambda)L_{n-1,a}(\lambda)\ldots L_{1,a}(\lambda)
=\left(\begin{array}{cc} A(\lambda) & B(\lambda) \\
                                     C(\lambda) & D(\lambda)
             \end{array}\right)
\lab{Ta}
\ee
and analogously the fundamental monodromy matrix $T_f(\lambda)$.
Taking the trace of the monodromy matrices over the auxiliary space we get two
operators, the auxiliary and fundamental transfer matrices,
\be
\Lambda(\lambda)=\tr_a\ T_a(\lambda), \qqqquad
\tau(\lambda)=\tr_f\ T_f(\lambda)
\lab{tr}
\ee
which act in the full quantum space of the model and commute with
each other for different values of the spectral parameters \ci{ABA}
\be
 [\tau(\lambda),\Lambda(\mu)]
=[\tau(\lambda),\tau(\mu)]
=[\Lambda(\lambda),\Lambda(\mu)]=0
\lab{tr-tr}
\ee
Differentiating the both sides of this relation with respect to
the spectral parameters $\lambda$ and $\mu$ and putting $\lambda=\mu=0$
we will get a family of mutually commuting conservation laws of the model.
Moreover, the fundamental transfer matrix $\tau(\lambda)$ contains
the local integrals of motion including the hamiltonian of the model, while
the operator $\Lambda(\lambda)$ allows one to construct
their eigenstates by means of the Bethe ansatz. The explicit form of the local
integrals of motion is given by \ci{TTF}
\be
I_k=\left.\frac1{i}\frac{d^k}{d\lambda^k}
    \log \tau(\lambda)\right\vert_{{}_{\lambda=0}}
   =\left.\frac1{i}\frac{d^k}{d\lambda^k}
    \log \tr_f\ T_f(\lambda)\right\vert_{{}_{\lambda=0}}
, \qquad k=1,2,\ldots
\lab{Ik}
\ee
where the $k-$th operator describes the interaction between $k+1$ nearest
neighbors on the lattice. In particular, the operator
$I_1=H=\sum_{k=1}^n H_{k,k+1}$ coincides with the hamiltonian of the model.

To find the explicit form of the operator $\Lambda(\lambda)$
we substitute the spin operators \re{S} for $s=0$ into the standard form
of the auxiliary Lax operator \re{Lax} and represent the result as a matrix in
$\IC^2$
\be
L_{k,a}^{(s=0)}(\lambda)= \lambda I + i v_k\otimes \bar u_k\
\partial_k
\lab{La0}
\ee
where $\partial_k\equiv\partder{z_k}$ and
$v_k=\bin{1}{z_k}$ and $\bar u_k=(z_k,\ -1)$ are vectors
in the auxiliary space $\IC^2$.
After substitution of \re{La0} into \re{Ta} we use the relation
$\bar u_k v_j = (z_k-z_j)$
to get
\be
\Lambda_{s=0}(\lambda)=2\lambda^n - S(S+1)\lambda^{n-2} + Q_3 \lambda^{n-3}
+\ldots + Q_n
\lab{Q_k}
\ee
where the operator $S$ is defined as
$$
S(S+1)=-\sum_{k>j}z_{kj}^2\partial_k\partial_j=\vec S \,\vec S=
S^3S^3+\frac12\lr{S^+S^-+S^-S^+}
$$
and $\vec S$ is the total spin of lattice.
The operators $Q_k$ for $k=3,\ldots,n$ are given by
$$
Q_k =  \sum_{n\ge i_1>i_2>\cdots>i_k\ge 1}
      i^k z_{i_1i_2} z_{i_2i_3}\ldots z_{i_ki_1}
      \partial_{i_1}\partial_{i_2}\ldots\partial_{i_k}
$$
and we recognize the operator $Q_n$ as the operator $A$ defined in \re{A}.
Notice, that in the expansion of $\Lambda_{s=0}(\lambda)$ the term
with $\lambda^{n-1}$ is absent due to the orthogonality of vectors $v_k$ and
$\bar u_k$. It follows immediately from \re{tr-tr} that the integrals
of motions $I_k$ and $Q_j$ are mutually commuting operators
$$
[I_k,Q_j]=[Q_k,Q_j]=[I_k,I_j]=0
$$
with $I_1=H$ being the holomorphic hamiltonian.

To diagonalize these operators one may try to apply the algebraic
Bethe ansatz \ci{ABA}. However, the attempt to use the Bethe ansatz for $s=0$
fails from the very beginning. The algebraic Bethe ansatz is based on
the existence of the highest weights in each cite of the lattice.
Trying to define them, we find that for $s=0$ the equations
$S_k^+\ket{\omega_k}=0$ and $S_k^3\ket{\omega}=s\ket{\omega}$ have
a trivial solution $\ket{\omega_n}=const$ which is annihilated by the
spin operator $\vec S$.
The reason why it happens is the following. As we will show in sect.4
the spin operators form for $s=0$ the
representation $t^{0,2}(g)$ of the principal series of the
noncompact group $SL(2,\IC)$.
This is a unitary infinite dimensional representation \ci{ZS}
which does not contain a highest weight,
in contrast with unitary finite dimensional representations of
$SU(2)$. Thus, to diagonalize the hamiltonian of the model one
has to generalize the algebraic Bethe ansatz to the case of
noncompact groups. Nevertheless, there is a trick which allows
one to apply the algebraic Bethe ansatz to find special eigenstates
of the model.

\section{Algebraic Bethe ansatz for spin $s=-1$}

There is
one-to-one correspondence between the XXX models of spin $s=0$ and $s=-1$
based on the following relation between the Lax operators \re{Lax}
in both models
$$
L_{k,a}^{(s=-1)}(\lambda)= \lambda I + i \partial_k\ v_k\otimes \bar u_k
= \lr{L_{k,a}^{(s=0)}(\lambda)}^T
= P_k\ L_{k,a}^{(s=0)}(\lambda)\ P_k^{-1}
$$
where operator $T$ denotes similarity transformation.
The fundamental Lax operators \re{Lax}
satisfy the same relation. From this property and the definition \re{Ta}
we find the relation between the monodromy matrices of the models
$$
T_a^{(s=-1)}(\lambda)=\lr{T_a^{(s=0)}(\lambda)}^T
=P_1P_2\ldots P_n\ T_a^{(s=0)}(\lambda)\ (P_1P_2\ldots P_n)^{-1}
$$
and analogously for $T_f^{(s=-1)}(\lambda)$. It is obvious from \re{tr}
that the same relations are valid between the transfer matrices but in
this case one can use the fact that they commute with the operator
$Q_n\equiv A$ defined in \re{A} to write
\be
\Lambda_{s=-1}(\lambda) = \Lambda_{s=0}^T(\lambda)
= (z_{12}z_{23}\ldots z_{n1})^{-1}\ \Lambda_{s=0}(\lambda)\
  z_{12}z_{23}\ldots z_{n1}
\lab{rel1}
\ee
and analogously for $\tau_{s=-1}(\lambda)$.
Thus, the transfer matrices in the XXX models of spin $s=0$ and $s=-1$ have
the same eigenvalues and their eigenstates are related as follows
\be
\ket{\varphi(z_i;z_0)}
=z_{12}z_{23}\ldots z_{n1}\ket{\hat\varphi(z_i;z_0)}
\lab{rel}
\ee
As follows from \re{H12} the same relation \re{rel1}
holds between the hamiltonians of both models.

The reason why we included the XXX model of spin $s=-1$ into consideration
is that one is able to apply the algebraic Bethe ansatz for spin
$s=-1$.%
\footnote{However, as is shown in sect.4, the algebraic Bethe ansatz
          gives us only special set of the eigenstates}
In contrast with spin $s=0$, it is possible to find a nontrivial
highest weight for spin $s=-1$.
Indeed, by substituting the explicit form of the spin
operator \re{S} for $s=-1$ into the equations
$S_k^+\ket{\omega_k}=0$ and $S_k^3\ket{\omega_k}=-\ket{\omega_k}$
we find that it has the nontrivial solution
$
\ket{\omega_k} = 1/{z_k^2}
$
which allows one to construct the pseudovacuum state as
$$
\ket{\Omega} = \frac1{z_1^2 z_2^2\ldots z_n^2}
$$
To define the Bethe states one has to use the $B$ operator
for spin $s=-1$ defined in \re{Ta}.
Then the Bethe states are given by
\be
\ket{\hat\varphi_l(\{\lambda\})}=B(\lambda_1)B(\lambda_2)
\ldots B(\lambda_l) \frac1{z_1^2 z_2^2\ldots z_n^2}
\lab{state}
\ee
where the parameters $\{\lambda\}=(\lambda_1,\ldots,\lambda_l)$ are solutions
of the Bethe equation for $s=-1$
\be
\lr{\frac{\lambda_k-i}{\lambda_k+i}}^n
=\prod_{j=1,\ j\neq k}^n \frac{\lambda_k-\lambda_j+i}{\lambda_k-\lambda_j-i}
\lab{B-eq}
\ee
and the operator $B$ is defined through \re{Ta} as an
element of the auxiliary monodromy matrix
corresponding to spin $s=-1$.
The Bethe states \re{state} depend on a positive integer number $l$ which
is fixed by the additional constraint \re{con1}. Indeed, using
\re{rel} it can be
easily shown that the transformation properties \re{con1} of an eigenstate
of the XXX hamiltonian for $s=0$ are equivalent to the following conditions
for the eigenstates of the XXX hamiltonian for spin $s=-1$
\be
S^3 \ket{\hat\varphi(z_i;0)} = - h \ket{\hat\varphi(z_i;0)}\,,\qquad
S^+ \ket{\hat\varphi(z_i;0)} = 0\,,\qquad
\ket{\hat\varphi(z_i;z_0)}=\ket{\hat\varphi(z_i-z_0;0)}
\lab{const}
\ee
where $\vec S$ is the total spin of the lattice for $s=-1$.
Comparing these constraints  with the properties of
the Bethe states, $S^3\ket{\hat\varphi}=-(n+l)\ket{\hat\varphi}$
and $S^+\ket{\hat\varphi}=0$, we find that
$$
n+l=h\,,\qqqquad l=0,\,1,\,\ldots
$$
where $h$ was defined in \re{h} and $n$ is the number of sites, or
equivalently,
the number of reggeized gluons in the $t-$channel. This relation means that
the algebraic Bethe ansatz allows us to construct a very special class of
eigenstates corresponding to {\it positive integer values\/} of the conformal
weight $h$ of the $n-$gluon wave function. To find the remaining states for
arbitrary complex $h$ a generalization of the Bethe ansatz is required.

Using the relations \re{state} and \re{rel} we finally get the expression
for the Bethe states which diagonalize the original hamiltonian as follows
$$
\ket{\varphi_l(z_i;0;\{\lambda\})}=z_{12}z_{23}\ldots z_{n1}
B(\lambda_1)B(\lambda_2)
\ldots B(\lambda_l) \frac1{z_1^2 z_2^2\ldots z_n^2}
$$
The integrals of motion for the XXX model with spin $s=-1$
are related to those for XXX model with spin $s=0$ by equations
identical to equation \re{rel1} between the transfer matrices.
That is why they have the same eigenvalues in both models.
The explicit expressions for the eigenvalues of integrals of motions
for arbitrary spin $s$ have been found in algebraic Bethe ansatz \ci{TTF}
and we use these expressions for $s=-1$ to get
\be
I_k=\sum_{j=1}^l\frac1{i}\frac{d^k}{d\lambda^k_j}
\log\frac{\lambda_j+i}{\lambda_j-i}
\lab{h_k}
\ee
Note that this expression was found with the function $f(s,\lambda)$
that defines the normalization of the $R-$matrix
in \re{Rf} chosen to be equal to one.
It follows from \re{Ik}, that
for arbitrary $f(s,\lambda)$ the expression \re{h_k}
gets an additional trivial contribution
$-in\frac{d^k}{d\lambda^k}f(s,\lambda)|_{\lambda=0}$.
Given a solution to the Bethe equation \re{B-eq} this relation
yields the spectrum of
local integrals of motion in the original model with hamiltonian $H$. In
particular, the eigenvalues of the hamiltonian are equal to
$$
\varepsilon_n\equiv I_1(\lambda)=-2\sum_{k=1}^l \frac1{\lambda^2_k+1}
$$
where $\{\lambda\}$ obey the Bethe equation \re{B-eq}
for a fixed number of reggeized gluons $n$ and $s=-1$.
Thus, for special values of the conformal weights $h$ the algebraic
Bethe ansatz constructed in this subsection yields the wave function
of the model \re{H-N} and the corresponding eigenvalue of the QCD hamiltonian.

\section{Eigenstates as highest weights of
representations of $SL(2,\IC)$}

So far we have considered only the properties of the holomorphic
hamiltonian in \re{H-N}. If we know its eigenstates we can generalize
them for the antiholomorphic hamiltonian in \re{H-N} and construct the
solution of the system \re{H-N} as the product of holomorphic and
antiholomorphic eigenstates
$$
\chi(\{z_i,\bar z_i\};z_0,\bar z_0)=
{\varphi(z_i;z_0)}
{\bar\varphi(\bar z_i;\bar z_0)}
$$
Let us combine the holomorphic and antiholomorphic hamiltonian to
define the following operator acting on both $z$ and $\bar z$ coordinates of
particles:
$$
\CH = H_n^T + \bar H_n
$$
where similarity transformation was defined in \re{rel1}.
This hamiltonian has the following properties.
As we have seen in sect.3, the hamiltonian $H_n$ describes the XXX
magnet with the generalized spin $s=0$ while the hamiltonian
$H_n^T$ corresponds to the XXX magnet with spin $s=-1$.
Recall that the hamiltonians $H_n$ and $H_n^T$ are related to each
other by an equation similar to \re{rel1} and they have the same spectrum of
eigenvalues and the corresponding eigenstates satisfy relation \re{rel}.
Hence, for fixed $\varepsilon_n$ and $\bar\varepsilon_n$ in
\re{H-N} the hamiltonian $\CH$ has the eigenvalue
$\varepsilon_n+\bar\varepsilon_n$ and the corresponding eigenstate is
\be
\hat\chi(\{z_i,\bar z_i\};z_0,\bar z_0)=
{\hat\varphi(z_i;z_0)}
{\bar\varphi(\bar z_i;\bar z_0)}=(z_{12}z_{23}\ldots z_{n1})^{-1}
\chi(\{z_i,\bar z_i\};z_0,\bar z_0)
\lab{phi-n}
\ee
Thus, there is one-to-one correspondence between this function and
the eigenstate $\chi$. We notice that the hamiltonian $\CH$ is
selfadjoint and its eigestates $\hat\chi$ are orthogonal to each other.
The reason
why we consider the functions $\hat\chi(\{z_i,\bar z_i\};z_0,\bar z_0)$
is that these functions are the highest weight
of the irreducible infinite dimensional unitary representation of the
principal series of the $SL(2,\IC)$ group.

To show this
we recall that the unitary representation of the principal series,
$t^{\rho,m}(g)$, is realized on the space of square integrable
functions $f(z,\bar z)$ according to
$$
t^{\rho,m}(g) f(z,\bar z)=(cz+d)^{-k/2+i\rho/2-1}
(\bar c\bar z +\bar d)^{k/2+i\rho/2-1}f\lr{\frac{az+b}{cz+d},
\frac{\bar a\bar z+\bar b}{\bar c\bar z+\bar d}}
$$
where $k$ is an integer, $\rho$ is a real number and the scalar
product in this space is given by
$$
\langle f_1 | f_2 \rangle = \int dz d\bar z f_1(z,\bar z) f_2^*(z,\bar z)
$$
In this representation
the group generators can be realized as holomorphic spin operators \re{S}
with $s=-1/2-k/4+i\rho/4$ and antiholomorphic spin operators
with $\bar s=-1/2+k/4+i\rho/2$.

For the one-dimensional model with the hamiltonian $\CH$
we have in each site $k$ of the lattice
holomorphic spin operators
$S^\pm_k$ and $S^3_k$ defined by \re{S} for $s=-1$ and antiholomorphic
spin operators $\bar S^\pm_k$ and $\bar S^3_k$ given by expressions \re{S}
for $s=0$ with the replacement of $z$ by $\bar z$. These six operators
commute with the hamiltonian $\CH$ (and not with
the QCD hamiltonian $H_n+\bar H_n$!) and
form the representation $t^{0,2}$ of the principal series of the $SL(2,\IC)$
group which we identify as a local quantum space in the $k-$th site. The full
spin operators of the lattice $S^\alpha=\sum_{k=1}^n S^\alpha_k$
and $\bar S^\alpha=\sum_{k=1}^n\bar S^\alpha_k$ act on the tensor
product of these representations which is reducible with respect to
$SL(2,\IC)$ and which can be decomposed into the direct sum of the
irreducible unitary representations of the principal series as follows.
It is well known \ci{ZS} that the product of two infinite dimensional
unitary representations of the principal series,
$t^{\rho_1,2m_1}$ and $t^{\rho_2,2m_2}$, where
$m_1$ and $m_2$ are integer and $\rho_1$ and $\rho_2$ are real,
is decomposed into the direct sum of the irreducible representations of the
same principal series $t^{\rho_3,2m_3}$ with $m_3$ being integer.
Applying this result to the tensor product of $n$ copies of
$t^{0,2}$ corresponding to all site of the lattice we get
$$
\underbrace{
t^{0,2}\otimes t^{0,2}\otimes \ldots \otimes t^{0,2}
}_n = \bigoplus_{\rho,m} t^{\rho,2m}
$$
that is the total quantum space of the lattice is decomposed into
the sum of infinite dimensional unitary representations of the principal
series of $SL(2,\IC)$ with real $\rho$ and integer $m$. In principle,
one may try to evaluate the multiplicities and the Clebsch-Gordan
coefficients for each $t^{\rho,2m}$ in
this decomposition using the standard formulas \ci{ZS}.
Thus, the eigenstates $\hat\chi(\{z_i,\bar z_i\};z_0,\bar z_0)$
of the hamiltonian $\CH$ belong to
the representations of the principal series $t^{\rho,2m}$. However,
there are additional constraints on the functions
$\hat\chi(\{z_i,\bar z_i\};z_0,\bar z_0)$ imposed by the transformation
properties \re{con1}, or equivalently by the conditions \re{const}
for holomorphic states
and
similar relations for antiholomorphic states. Using the relation \re{phi-n}
we represent these conditions in the following form
$$
S^3\ket{\hat\chi}
=-h\ket{\hat\chi},\qqquad
S^+\ket{\hat\chi}=0,\qqquad
\bar S^3\ket{\hat\chi}
=-\bar h\ket{\hat\chi},\qqquad
\bar S^+\ket{\hat\chi}=0
$$
where $\hat\chi\equiv\hat\chi(\{z_i,\bar z_i\};z_0=0,\bar z_0=0)$ and
the conformal weights $h$ and $\bar h$ were defined in \re{h}.
Using the explicit form of $h$ and $\bar h$ we conclude that the
eigenstates of the hamiltonian $\CH$ which obey these additional conditions
are the highest weights of the irreducible representation
$t^{4\nu,2m}$ of the principal series of the $SL(2,\IC)$.
Thus, for arbitrary complex conformal weights $h$ and $\bar h$ in \re{h}
there might exist the eigenstate of the model.
On the other hand, as was shown in sect.3, the algebraic Bethe ansatz is
applicable only for integer $h$ and it cannot be generalized even for
noninteger conformal weights.

\section{Generalized Bethe ansatz}

To find the eigenstates corresponding to arbitrary complex values of the
conformal weights one could use the method of the $Q-$operator \ci{Baxter}
proposed by Baxter. Using the relations between the XXX models of spins
$s=0$ and $s=-1$
we restrict ourselves only by consideration of the model with
$s=-1$. In this case, there exists an operator $Q(\lambda)$
which acts on the full quantum space of the lattice and obeys the
Baxter equation
\be
\Lambda(\lambda)Q(\lambda)=(\lambda-i)^n Q(\lambda-i)
                          +(\lambda+i)^n Q(\lambda+i)\,,
\qquad
[Q(\lambda),\Lambda(\mu)]=[Q(\lambda),Q(\mu)]=0
\lab{Q-s}
\ee
Thus, the operators $\Lambda$ and $Q$ can be diagonalized simultaneously.
Their corresponding eigenvalues are $c-$number valued functions of the
spectral parameter $\lambda$ which obey the same equation \re{Q-s}.
That is why we are using below the same notations $\Lambda(\lambda)$
and $Q(\lambda)$ for the eigenvalues of these operators.
Solving this functional equation and using the
analyticity properties
of $\Lambda(\lambda)$ as a function of $\lambda$, one is able to find the
eigenvalues of the auxiliary transfer matrix $\Lambda(\lambda)$ and the
operator $Q(\lambda)$.
The standard way of solving the Baxter equation consists of choosing
eigenvalues $Q(\lambda)$ to depend polynomially on the spectral parameter
$\lambda$
\be
Q(\lambda)=\mbox{const.}\prod_{k=1}^l (\lambda-\lambda_k)
\lab{Q-B}
\ee
where $\{\lambda\}=\{\lambda_1,\ldots,\lambda_l\}$ are arbitrary complex
numbers and $l$ is positive integer.
The parameters $\{\lambda\}$ are found by imposing analyticity conditions
for the function $\Lambda(\lambda)$ at $\lambda=\lambda_k$. Namely,
substituting $\lambda=\lambda_k$ into the both sides of the Baxter
equation we get an equation for the parameters $\{\lambda\}$ which coincides
with the Bethe equation \re{B-eq}. Thus, the algebraic Bethe ansatz
corresponds to the special case of the eigenvalues of the $Q-$operator
as finite polynomials. To go beyond the algebraic Bethe ansatz one has
to look for solutions of the Baxter equation among
the functions more general than polynomials. The simplest possibility
which is realized in the Toda model is to search for the solution
among the entire functions having {\it infinite\/} number of zeros,
$l\to\infty$.

Using the definition \re{Q_k} we find that the function
$\Lambda(\lambda)$ is a polynomial of power $n$ in $\lambda$ with
the coefficients given by the operators $S(S+1)$ and $Q_k$.
This suggests the following way to solve the Baxter equation.
First, one substitutes the general form of the function
$\Lambda(\lambda)$ into the l.h.s. of the Baxter equation,
replacing all operators by their unknown eigenvalues,
$S$ and $Q_k$ $(k=3,\ldots,n)$.
Second, one solves the functional equation \re{Q-s} and finds the
solution for $Q(\lambda)$ which depends on these
quantum numbers. Third, imposing the condition that $Q(\lambda)$ is
an entire function of
$\lambda$ one determines the possible values of $S$ and $Q_k$ and substitutes
them into \re{Q_k} to find the eigenvalue of the auxiliary transfer matrix.

Thus, for any fixed allowed set of the quantum numbers $\{S,Q_k\}$ we will
find the corresponding function $Q(\lambda)$ satisfying the Baxter equation.
It is important to notice that among these functions there
are very special ones \re{Q-B} corresponding to the Bethe ansatz solution of
the XXX model of spin $s=-1$. Moreover, in this
particular case we know the explicit form of the eigenstates
\re{state} and eigenvalues \re{h_k} of the local
integrals of motion. All these expressions, including the $Q-$function
\re{Q-B}, explicitly depend on the parameters $\{\lambda_k\}$ which are
solutions of the Bethe equation \re{B-eq}.
The natural question appears: is it possible to
express both the eigenstates \re{state} and the eigenvalues
\re{h_k} in terms of $Q(\lambda)$ in
such a way, that all their dependence on $\{\lambda_k\}$ will be contained
inside the $Q-$function? In this case, one will get a unique possibility
to perform an ``analytical continuation'' of the results found within the
framework of the algebraic Bethe ansatz by replacing the special values
\re{Q-B} of the $Q-$function by a general solution of the Baxter equation.

Let us express the Bethe states \re{state} in terms of the $Q-$function
defined in \re{Q-B}. Substituting the explicit form of the auxiliary Lax
operator into the definitions \re{Ta} one finds that for $s=-1$ the
operator $B(\lambda)$ is a polynomial of power $n-1$ in $\lambda$
which can be represented in the following form
\be
B(\lambda) = iS^- (\lambda-x_1)(\lambda-x_2)\ldots (\lambda-x_{n-1})
\lab{x_k}
\ee
where $x_1,\ldots,x_{n-1}$ are operator zeros of $B(\lambda)$. The precise
meaning of $x_k$ as roots of the operator polynomial $B(\lambda)$ was
given in the framework of the functional Bethe ansatz \ci{Skl85,Skl92}.
These operators have the following properties
$$
[S^- , x_k] = [S^3 , x_k] = [x_k , x_j ] = 0
$$
which allows us not to worry about their ordering in \re{x_k}.
Now we may rewrite the Bethe states \re{state} in terms of the operators $x_k$
$$
\ket{\hat\phi_l} =  \prod_{j=1}^l iS^- \prod_{k=1}^{n-1} (\lambda_j-x_k)
\ket{\Omega} 
= (iS^-)^l \prod_{k=1}^{n-1} \prod_{j=1}^l (\lambda_j-x_k)
\ket{\Omega} 
$$
where $\lambda_j$ satisfies the Bethe equation \re{B-eq} for $s=-1$ and
fixed $l$ and
$n$. Now we immediately recognize that the product over $j$ coincides with
the expression \re{Q-B} for the function $Q(\lambda)$ defined for
$\lambda=x_k$.
Notice that the $Q-$operator doesn't commute with the operators $x_k$ and
the notation $Q(x_k)$ means the eigenvalue of the $Q-$operator defined
to be evaluated for operator value of the spectral parameter.
Thus, the Bethe states can be expressed in terms of the eigenvalues of
the $Q-$operator as follows
\be
\ket{\hat\phi} = (iS^-)^{h-n} Q(x_1)Q(x_2)\ldots Q(x_{n-1})
\frac1{z_1^2z_2^2\ldots z_n^2}
\lab{STATE}
\ee
where $h$ is the conformal weight defined in \re{h}.
The remarkable property of this expression is that it does not depend
explicitly on the parameters $\lambda_k$. The operators $S^-$ and $x_k$
are determined by the properties of the model and are the same for
all eigenstates. Different eigenstates $\ket{\hat\phi_l}$ are parameterized
by different solutions $Q(\lambda)$ of the Baxter equation.
After being written in this form, the relations \re{STATE} admit a natural
analytic continuation to arbitrary complex values of $S$, or equivalently
$h$. Moreover, these states satisfy the additional conditions \re{const}
which follow from the analytical continuation of the analogous relations
for the Bethe states.

Once we know the eigenvalues of the $Q-$operator
the eigenvalues of the auxiliary transfer matrix can be easily found
from \re{Q-s}.
To find the eigenvalues of the local integrals of motion we use
their expressions found by the algebraic Bethe ansatz method, rewrite
them in terms of the $Q-$functions and perform analytical continuation.
Using equation \re{h_k} we represent the eigenvalues of the operator
$I_k$ for $s=-1$ in the following form
$$
I_k=\frac1{i}\sum_{j=1}^l\frac{d^k}{d\lambda_j^k}
    \log\frac{\lambda_j+i}{\lambda_j-i}
   =
    \frac1{i}\frac{d^k}{d\lambda^k}\log\prod_{j=1}^l
    \left.
    \frac{\lambda_j+i+\lambda}{\lambda_j-i+\lambda}
    \right\vert_{{}_{\lambda=0}}
$$
and in terms of $Q-$function \re{Q-B} this expression looks like
\be
I_k=\left.
    \frac1{i}\frac{d^k}{d\lambda^k}\log
    \frac{Q(-\lambda-i)}{Q(-\lambda+i)}
    \right\vert_{{}_{\lambda=0}}
\lab{IK}
\ee
Although this relation was found as a relation between eigenvalues of
the operators $I_k$ and $Q$ it can be extended to be an operator relation
because both operators have the same eigenstates.

\section{Solution of the Baxter equation}

To find the eigenstates and eigenvalues of the integrals of motion one has
to solve the Baxter equation for $s=-1$ and substitute the solution for
the function $Q(\lambda)$ into the relations \re{STATE}
and \re{IK}.
Using the general
form of the function $\Lambda(\lambda)$ in terms of quantum numbers
$\{S,Q_k\}$ we get the following functional equation for the
function $Q(\lambda)$
\be
(2\lambda^n-h(h-1)\lambda^{n-2}+Q_3\lambda^{n-3}+\ldots+Q_{n})Q(\lambda)
=(\lambda+i)^n Q(\lambda+i)+(\lambda-i)^n Q(\lambda-i)
\lab{Bax}
\ee
where we used the relations \re{const} and \re{Q_k} to
identify the eigenvalue of the operator $S$ with the conformal weight $h$
of the eigenstates defined in \re{h}. In this equation $h$ and $Q_k$ are
independent parameters whose values are restricted by analyticity properties
of the solution $Q(\lambda)$.

The Baxter equation \re{Bax} depends on the set of quantum numbers
$\{h,Q_k\}$
and obeys the following property. It is invariant under the replacement
\be
h\to 1-h
\lab{1-h}
\ee
Hence, if the function $Q(\lambda;h,\{Q_k\})$ is a solution of the Baxter
equation then so is $Q(\lambda;1-h,\{Q_k\})$.
Combined with \re{STATE}, this property allows us to relate the
eigenstates of the model with the conformal weights $h$ and $1-h$.
Thus, trying to solve the Baxter equation \re{Bax}
we use the symmetry \re{1-h} to restrict the possible values of the
conformal weights to the fundamental region
\be
\mbox{Re } h \ge \frac12
\lab{h>1/2}
\ee
or in terms of the parameters $m$ defined in \re{h}, $m\ge 0$. It is
interesting to note that the Bethe ansatz solution \re{Q-B} for $Q(\lambda)$
corresponds to integer positive $h$ and the symmetry \re{1-h} allows us to
generalize the results of the sect.3 to arbitrary integer $h$ by putting
$l=|h-\frac12|-n+\frac12$.
This is the simplest example of the analytical continuation of results
obtained by means of the Bethe ansatz.

In the limit of large $\lambda$ the Baxter equation can be rewritten as
$$
-h(h-1) \lambda^{n-2}Q(\lambda)= (\lambda+i)^n Q(\lambda+i)
                 + (\lambda-i)^n Q(\lambda-i)
                 -  2\lambda^n Q(\lambda)
                 \sim
                 i^2 \frac{d^2}{d\lambda^2}\lr{\lambda^nQ(\lambda)}
$$
where we neglected nonleading terms in both sides of the equation.
Solving the resulting second order differential equation we get the
general large$-\lambda$ asymptotic behavior of the solutions of the
Baxter equation in the fundamental region \re{h>1/2}
\be
Q(\lambda) \stackrel{\lambda\to\infty}{\sim} \lambda^{h-n}\equiv\lambda^l
\lab{Q-as}
\ee
Using the asymptotics \re{Q-as} we recognize the special
role of integer positive
values of the conformal weights $h$. The analyticity of $Q(\lambda)$
as a function of $\lambda$ together with \re{Q-as} implies that for
$h-n\in\IZ_+$
the solution of the Baxter equation is polynomial of order $h-n$
in $\lambda$ which is in accordance with \re{Q-B}.

To solve the Baxter equation \re{Bax} for arbitrary $\lambda$ we perform
the Mellin transformation of the function $Q(\lambda)$
\be
Q(\lambda)=\int_0^\infty d\omega \ \omega^{i\lambda-1} Q(\omega)
\lab{Q(x)}
\ee
After substitution of \re{Q(x)} into \re{Bax}, the original finite
differential equation for $Q(\lambda)$ is replaced by an ordinary $n-$th
order differential equation for $Q(\omega)$:
$$
\left[\frac{(\omega-1)^2}{\omega}\lr{i\omega\frac{d}{d\omega}}^n
-\sum_{k=0}^{n-2}Q_{n-k}\lr{i\omega\frac{d}{d\omega}}^k
\right]Q(\omega)=0
$$
where $Q_2\equiv -h(h-1)$. A simple analysis shows that the solution of this
equation has singularities at $\omega=1$. Indeed, in the large $\lambda$ limit
the function \re{Q(x)} gets its leading contribution from the integration at
the vicinity of the point $\omega=1$ and the asymptotics \re{Q-as} imply that
$$
Q(\omega) \stackrel{\omega\to 1}{\sim} (1-\omega)^{-1-h+n}
$$
where $h$ takes values in the fundamental region \re{h>1/2}. Thus, for the
expression \re{Q(x)} to be well defined one has to fix the prescription
for the integration of the $\omega=1$ singularity. Since the integration in
\re{Q(x)} is performed along the real positive axis in the complex
$\omega-$plane one can deform the integration path at the point $\omega=1$
to encircle the singularity either in lower or upper half plane. Moreover,
rotating both integration paths as $\omega\to\e^{i\pi}\omega$ and
$\omega\to\e^{-i\pi}\omega$, respectively, we avoid the singular point
$\omega=1$ and get the following expression for the $Q-$function
$$
Q(\lambda)=\lr{C_+\e^{\pi\lambda}+C_-\e^{-\pi\lambda}}
\int_0^\infty d\omega \ \omega^{i\lambda-1} Q(-\omega)
$$
where $C_+$ and $C_-$ are arbitrary constants. However the value of the
constants is fixed by the condition that for arbitrary complex $\lambda$
the integral should be convergent at $\omega=0$ and $\omega=\infty$
\be
C_+=-C_-=\frac12
\lab{C+-}
\ee
where the numerical value can be arbitrary.
Changing the integration
variable $z=1/(1-\omega)$ we obtain
\be
Q(\lambda)=\sinh(\pi\lambda)
\int_0^1 dz \ (1-z)^{i\lambda-1} z^{-i\lambda-1}Q(z)
\lab{Q(z)}
\ee
where the function $Q(z)$ obeys the following equation
$$
\left[\lr{-iz(1-z)\frac{d}{dz}}^n
+z(1-z)\sum_{k=0}^{n-2}Q_{n-k}\lr{-iz(1-z)\frac{d}{dz}}^k
\right]Q(z)=0
$$
Let us consider the solutions of this equation in the special case
$n=2$ corresponding to the compound state of two reggeized gluons -
the BFKL pomeron.

For $n=2$ the equation for the function $Q(z)$ has the following form
$$
z(1-z)Q''(z)+(1-2z)Q'(z)+h(h-1)Q(z)=0
$$
where prime denotes differentiation with respect to $z$. The equation
is invariant under the replacement $z\to 1-z$ and its general solution
is well known \ci{hyper} to be a linear combination of the hypergeometric
functions
$$
Q(z)=c_1 F(h,1-h;1;z) - c_2 F(h,1-h;1;1-z)
$$
where $c_1$ and $c_2$ are arbitrary constants. After substitution of this
expression into \re{Q(z)} we find the $Q-$function as
\be
Q(\lambda)=
           c_1 Q_0(\lambda) + c_2 Q_0(-\lambda)
\lab{Q-ss}
\ee
where $Q_0(\lambda)$ denotes the function
$$
Q_0(\lambda)=\sinh(\pi\lambda)
\int_0^1 dz \ (1-z)^{i\lambda-1} z^{-i\lambda-1}F(h,1-h;1;z)
$$
We use the integral representation for the hypergeometric function
\ci{hyper} in order to express the solution in the following form
\be
Q_0(\lambda)={\sinh(\pi\lambda)}\ {\sin(\pi h)}
\int_0^1 dz \ (1-z)^{i\lambda-1} z^{-i\lambda-1}
\int_0^1 dt \ t^{h-1} (1-t)^{-h} (1-tz)^{h-1}
\lab{Q0-i}
\ee
This expression for $Q_0(\lambda)$ can be represented as a
double Pochhammer contour integral
$$
Q_0(\lambda)=\frac{i}4\oint_P dz\,z^{-i\lambda-1} (z-1)^{i\lambda-1}
\oint_P dt\,t^{h-1} (t-1)^{-h} (1-tz)^{h-1}
$$
where the contour $P$ incloses the singular points $0$ and $1$ in complex
$z$ and $t$ planes.
Taking \re{Q0-i} and expanding the last factor in the integrand in
powers of $zt$ we perform the integration over $z$ and get the following
factor
$
\Gamma(i\lambda)\Gamma(1-i\lambda)=-\frac{i\pi}{\sinh(\pi\lambda)}
$
which generates the singularities of the function $Q(\lambda)$ at
$\lambda=i\IZ$. They originate from the ``dangerous'' points $z=0$ and $z=1$
in the integral. To preserve the analyticity of the $Q-$function
we use the possibility to choose arbitrary values of the constants $C_+$
and $C_-$ to put them equal to \re{C+-}. This leads to the appearance of
the factor $\sinh(\pi\lambda)$ in \re{Q0-i} which compensates the
singularities of the integral.
Finally, we get the representation for $Q_0$ as an infinite sum
\be
Q_0(\lambda)=
\sum_{k=1}^\infty (-)^k\frac{k}{(k!)^3}\frac{\Gamma(h+k)}{\Gamma(h-k)}
\frac{\Gamma(k-i\lambda)}{\Gamma(1-i\lambda)}
=\left. h(h-1)\ {}_3F_2(1+h,2-h,1-i\lambda;2,2;z)\right|_{z\to 1}
\lab{Q0-s}
\ee
where ${}_3F_2$ is the generalized hypergeometric function \ci{hyper}.
This expression has the following properties.

In the $k-$th term of the sum the $\lambda-$dependence comes from
the ratio $\Gamma(k-i\lambda)/\Gamma(1-i\lambda)$ which is a
polynomial of order $k$ in $\lambda$. Hence, being expanded in
powers of $\lambda$ the function $Q_0(\lambda)$ turns out to be
an infinite series.
This seems to be in contradiction with the fact, that in the special
case of positive integer values of $l=h-2$ the Bethe ansatz gives us an
expression for the $Q-$function which is a polynomial of power $h-2$ in
$\lambda$. However, notice that in the representation \re{Q0-s} the factor
$1/\Gamma(h-k)$ ensures the truncation of the sum after the $k=(h-1)-$th term
for $h\ge 2$. Moreover, the expression \re{Q0-s} explicitly
obey the symmetry \re{1-h} which implies that
the series terminates also for negative integer $h$.

For arbitrary
noninteger $h$ the expression \re{Q0-s} is an infinite series in $\lambda$.
Using the properties \ci{hyper} of the function ${}_3F_2$ one finds that this
series converges only if $\mbox{Im}\ \lambda <0$. Then, the general form
of the solution \re{Q-ss} implies that the solution of the Baxter equation
$Q(\lambda)$ is given by the function $Q_0(\lambda)$ in the lower half plane
in $\lambda$ and by $Q_0(-\lambda)$ in the upper half plane. It is
interesting to note that for integer $h$ the function satisfies the
relation $Q_0(-\lambda)=(-1)^h Q_0(\lambda)$ which implies that
both terms in \re{Q-ss} are equivalent.

After substitution of \re{Q0-s} into \re{Q-ss} we get an expression
for $Q(\lambda)$ which is an entire function of $\lambda$ having an
infinite number of zeros. Only in the case of positive integer $h$ the
number of zeros is finite. It can be checked that the roots $\{\lambda_k\}$
do obey the Bethe equation for $n=2$. This is in accordance with general
conditions on the $Q-$function discussed in sect.5.

\section{Conclusions}

In this paper we found that the one-dimensional lattice model proposed
by Lipatov to describe high-energy scattering of hadrons in QCD is
completely integrable. Applying the quantum inverse scattering method we
identified the Lipatov model as the generalized one-dimensional XXX chain
of spin $s=0$. We found the family of local integrals of motion
and developed the generalized Bethe ansatz for their diagonalization.
The corresponding eigenstates and eigenvalues, \re{STATE} and \re{IK},
are expressed in terms of the $Q-$function which satisfies the Baxter
equation \re{Q-s} for spin $s=-1$. Solving this equation in the special
case of the lattice with $n=2$ sites we found the expressions \re{Q-ss}
and \re{Q0-s} for the $Q-$function. It turns out that after the substitution
of this function into \re{STATE} and \re{IK} we will get the expressions
for the wave function $\chi$ of the compound state of $n=2$ reggeized gluons
and the corresponding eigenvalue of the QCD hamiltonian which are identical
to that \ci{L86} for the BFKL pomeron. The details of the calculations and the
proof of the statements made in this paper will be published elsewhere.


\bigskip

We would like to thank V.Korepin and L.Takhtajan for helpful conversations.
L.D.F. is grateful to C.N.Yang for the hospitality at the Institute for
Theoretical Physics at Stony Brook which in particular made our collaboration
possible. G.P.K. is most grateful to J.de Boer and G.Sterman for
numerous stimulating discussions and for their support.
The work of G.P.K. is supported in part by the National Science Foundation
under grant PHY9309888.

\bb{99}
\bi{L94}
      L.N.Lipatov, {\it ``High energy asymptotics of multi-color QCD and
      exactly solvable lattice models''\/}, Padova preprint, DFPD/93/TH/70,
      October 1993
\bi{ABA}
      L.D. Faddeev,
      {\it ``Algebraic aspects of Bethe ansatz''\/},
      Stony Brook preprint, ITP-SB-94-11, Mar 1994; hep-th/9404013
      {\it ``The Bethe ansatz''\/}, Andrejewski lectures, Freie Univ.
      preprint, SFB-288-70, Jun 1993;
      {\it ``Lectures on quantum inverse scattering method''\/}
      in Nankai Lectures on Mathematical Physics,
      Integrable Systems, ed. by X.-C.Song, pp.23-70,
      Singapore: World Scientific, 1990.
\bi{KIB}
      V.E. Korepin, N.M.Bogoliubov and A.G. Izergin, {\it ``Quantum
      inverse scattering method and correlation functions''\/},
      Cambridge Univ. Press, 1993
\bi{L89}
      L.N.Lipatov, {\it ``Pomeron in quantum chromodynamics''\/},
      in ``Perturbative QCD'', pp.411--489,
      ed. A.H. Mueller (World Scientific, Singapore, 1989)
\bi{Cheng}
      H. Cheng, J. Dickinson, C.Y. Lo and K. Olaussen,
      {\it ``Unitarizing high-energy scattering amplitudes in field
      theories. 2. Yang-Mills theories.\/}, Nuovo Cim. Lett. 25 (1979) 175;
\\    {\it ``Diagrammatic derivation of the eikonal formula for high-energy
      scattering in Yang-Mills theory\/}, Phys. Rev. D23 (1981) 534-552.
\bi{book2}
      H. Cheng and T.T. Wu, {\it ``Expanding Protons: Scattering at
      High Energies''\/}, (MIT Press, Cambridge, Massachusetts, 1987).
\bi{Bar}
      J. Bartels, {\it ``High-energy behavior in a nonabelian gauge
      theory. 2. First corrections to $T_{n\to m}$ beyond the leading
      ln $s$ approximation''\/}, Nucl. Phys. B175 (1980) 365
\bi{KP}
      J. Kwiecinski and M. Praszalowicz,
      {\it ``Three gluon integral equation
      and odd C singlet Regge singularities in QCD''\/},
      Phys.Lett. B94 (1980) 413-416
\bi{L90}
      L.N.Lipatov, {\it ``Pomeron and odderon in QCD and a two dimensional
      conformal field theory''\/}, Phys. Lett. B251 (1990) 284
\bi{BFKL}
      E.A.Kuraev,  L.N. Lipatov and V.S. Fadin,
      {\it ``Multiregge processes in the Yang-Mills theory''\/},
      Sov.Phys.JETP, 44 (1976) 443-451;
\\     E.A.Kuraev,  L.N. Lipatov and V.S. Fadin,
      {\it ``The Pomeranchuk singularity in nonabelian gauge theories''\/},
      Sov.Phys.JETP 45 (1977) 199-204;
\\    Ya.Ya. Balitskii and L.N. Lipatov, {\it ``Pomeranchuk singularity in
      quantum chromodynamics''\/}, Sov.J.Nucl.Phys. 28 (1978) 822-829
\bi{L93}
      L.N.Lipatov, {\it ``High-energy asymptotics of multicolor QCD and
      two-dimensional conformal field theories''\/},
      Phys. Lett. B309 (1993) 394-396
\bi{L86}
      L.N.Lipatov, {\it ``The bare pomeron in quantum chromodynamics''\/},
      Sov.Phys.JETP 63 (1986) 904
\bi{KI}
      V.E. Korepin and A.G. Izergin, {\it ``Lattice model connected with
      nonlinear Schrodinger equation''\/},
      Sov. Phys. Doklady, 26 (1981) 653-654;
      {\it ``Lattice versions of quantum field theory models in two
      dimensions''\/}, Nucl. Phys. B205 (1982) 401-413
\bi{TTF}
      V.O.Tarasov, L.A.Takhtajan and L.D.Faddeev, {\it ``Local hamiltonians for
      integrable quantum models on a lattice''\/}, Theor. Math. Phys.
      57 (1983) 163-181
\bi{ZS} D.P. Zhelobenko and  A.I. Shtern, {\it ``Representations of Lie
      groups''\/} (in Russian), Nauka, Moscow, 1983, pp.211-220
\bi{Baxter}
      R.J. Baxter, {\it ``Exactly Solved Models in Statistical
      Mechanics''\/}, Academic Press, London, 1982
\bi{Skl85}
      E.K. Sklyanin,
      {\it ``The quantum Toda chain''\/},
      Lecture Notes in Physics (Springer) 226 (1985) 196-233;
\bi{Skl92}
      E.K. Sklyanin,
      {\it ``Quantum Inverse Scattering Method. Selected Topics''\/},
      in ``Quantum Group and Quantum Integrable Systems'' (Nankai
      Lectures in Mathematical Physics), ed. Mo-Lin Ge, Singapore:
      World Scientific, 1992, pp.63--97; hep-th/9211111
\bi{hyper} {\it ``Higher transcendental functions''\/} vol.1,
      Bateman manuscript project, ed. A.Erdelyi, McGraw-Hill, 1953
\eb
\end{document}